\documentclass[prl,preprint,showpacs]{revtex4}

\usepackage{amssymb,amsmath,bm}
\usepackage[final]{graphicx}

\newcommand{\modena}[0]{{INFM--$S^3$ and
Dipartimento di Fisica Universit\`a di Modena e Reggio Emilia, 
Via Campi 213/A, 41100 Modena, Italy}}

\newcommand{\graz}[0]{{
  Institut f\"ur Theoretische Physik,
  Karl--Franzens--Universit\"at Graz, Universit\"atsplatz 5,
  8010 Graz, Austria}}
  
\newcommand{\onehalf}[0]{\mbox{$\frac 1 2$}}

\begin{document}
\bibliographystyle{apsrev}

\title{High-finesse optical quantum gates for electron spins in artificial molecules}

\author{Filippo Troiani}\email{troiani@unimore.it}\affiliation{\modena}
\author{Ulrich Hohenester}\affiliation{\graz}
\author{Elisa Molinari}\affiliation{\modena}

\date{\today}

\begin{abstract}

A doped semiconductor double-quantum-dot molecule is proposed as a 
qubit realization. The quantum information is encoded in the electron 
spin, thus benefiting from the long relevant decoherence times; the 
enhanced flexibility of the molecular structure allows to map the spin 
degrees of freedom onto the orbital ones and vice versa, and opens the 
possibility for high-finesse (conditional and unconditional) quantum 
gates by means of stimulated Raman adiabatic passage.

\end{abstract}

\maketitle

 
Quantum bits or {\em qubits}\/ are the building block for future 
quantum computers \cite{bouwmeester:00,bennett:00}. The requirements 
for such quantum hardware are manifold: first, qubits should consist 
of at least two long-lived states, usually referred to as $0$ and 
$1$; second, it should be possible to modify the state of a single 
qubit unconditionally or dependent on the setting of a second qubit 
(one- and two-qubit quantum gates); finally, one should be able to 
measure the final qubit states. 
Evidently, the main challenge in identifying physical systems as 
qubits is to bridge between the two complementary requirements of 
long quantum memory and fast quantum gates: while the first point 
requires excitations well protected from environment, the latter one 
calls for strong and well-controllable interacting channels between 
the qubits and the external control.

To overcome this difficulty, in their seminal work Cirac and 
Zoller~\cite{cirac:95} proposed to pursue a mixed strategy in which 
the quantum information is stored in metastable atomic states and 
the light coupling to additional auxiliary states is used to perform 
the quantum gates. The recent progress in the fabrication and control 
of semiconductor quantum dots~\cite{hawrylak:98}, sometimes referred 
to as {\em `artificial atoms'},\/ suggested that similar schemes could 
be identified also in the technologically more promising solid 
state: indeed, in Refs.~\cite{troiani.prb:00,biolatti:00} optical 
excitations (excitons) in artificial atoms were proposed as qubits, 
with Coulomb interactions between the optically excited electrons and 
holes providing a means to perform conditional quantum gates. 
However, it was soon realized that the radiative lifetime of excitons 
($\sim$ns) is too short to comply with the exceptional requirements 
for quantum memory. 
Around the same time different work proposed spin of excess electrons 
as a viable quantum memory~\cite{loss:98,imamoglu:99}, and estimated 
lifetimes of the order of microseconds. Apparently, a combination of 
such long spin memory with ultrafast optical gating provides a likely 
candidate for the first proof-of-principle solid state quantum 
computer, in particular in view of the superb standards of presentday 
sample growth and coherent-carrier control. 
In turn, a diversity of strategies for such a mixed approach was put 
forward, e.g., based on cavity quantum 
electrodynamics~\cite{imamoglu:99}, charged excitons~\cite{pazy:02,cortez:02}, 
or RKKY interactions~\cite{piermarocchi:02b}. 
Yet, the shortcomings of these proposals are either lacking strategies 
for performing conditional or unconditional gates, or possible 
environment losses 
during gating, issues which might become limiting in view of the 
exceptional quantum-computation requirements.


It is the purpose of this paper to propose a quantum computation 
scheme based on long spin memory and ultrashort optical quantum 
gates {\em which does not suffer from radiative losses during 
gating}.\/ Most importantly and in contrast to all existing proposals 
based on electron spin in quantum dots, we consider a vertically 
coupled double dot ({\em `artificial molecule'}) as the building block for 
a single qubit. 
The quantum hardware then consists of laterally arranged quantum-dot 
molecules (e.g., through seeded growth~\cite{ishida:98}) 
which can be individually addressed through frequency selective laser 
pulses. Besides, in order for the qubits to be correctly defined, the 
interdot tunneling in the lateral directions has to be 
suppressed~\cite{loss:98}.
Our central observation concerns the fact that these 
artificial molecules host besides the spin-degenerate electron state 
(used as the qubit) further long-lived auxiliary states which can be 
exploited during gating {\em to map the information stored in the 
spin degrees of freedom onto the orbital ones and vice versa, and
optically switch on and off qubit-qubit interactions on an ultrashort timescale.}\/  As will be shown below, with this strategies it becomes possible to perform all quantum gates efficiently by means of stimulated Raman 
adiabatic passage (STIRAP)~\cite{bergmann:98} and to hereby suppress 
environment losses during gating.

 
{\em 1. Qubit identification.}\/ We start by considering two 
vertically-coupled quantum dots (see Fig.~1) inside a field-effect 
structure. The electric field in the growth direction has two 
consequences: first, it transfers a single excess electron from a 
nearby $n$-type reservoir to the artificial molecule, where further 
charging is suppressed because of the Coulomb blockade; second, it 
enhances the electron localization in one of the two dots (labeled 
as large, $L$, as compared to the small one, $S$). 
Although in the following we shall not be too specific about the 
details of this quantum-dot molecule (model calculations will be 
presented at the end), we assume that in presence of a uniform 
magnetic field along $x$ the electron eigenstates become a direct 
product of orbital and spin degrees of freedom, respectively:

\begin{subequations}\label{eq:states}
\begin{eqnarray}
  |0\rangle&=&|L\rangle\otimes|S_x=-\onehalf\rangle\\
  |1\rangle&=&|L\rangle\otimes|S_x=+\onehalf\rangle\\
  |2\rangle&=&|S\rangle\otimes|S_x=-\onehalf\rangle,
\end{eqnarray}
\end{subequations}
 
\noindent with $|L\rangle$ ($|S\rangle$) the orbital part associated 
to localization in the large (small) dot, and $ |S_x=\pm\onehalf
\rangle $ the spin part; states $|0\rangle$ and $|1\rangle$ will 
serve us for encoding the qubit, whereas state 
$ |2\rangle $ is an auxiliary state which will be used during gating.

Next, we introduce as a fourth (auxiliary) state $|3\rangle$, which 
allows optical coupling between the electrons states of 
Eq.~\eqref{eq:states}, the negatively charged exciton state 
$|X^-\rangle$ \cite{warburton:00,hartmann.prl:00,findeis.prb:01}, 
i.e., an electron-hole complex consisting of two electrons and a 
single hole; besides, we assume that in presence of the strong 
confinement along $z$ the hole acquires a well-defined symmetry 
because of the splitting of heavy- and light-hole states. 
Thus, in the qubit manipulation by means of external laser pulses 
the {\em light polarization}\/ allows to control the spin degrees 
of freedom (e.g., to individually address the 0--3 and 1--3 
transitions), whereas the {\em light frequency}\/ serves as a control 
for the orbital part~\cite{imamoglu:99} (e.g., to individually address 
0--3 and 2--3). The resulting optical selection rules are sketched 
in Fig.~1.


{\em 2. Quantum gates.}\/ As a major improvement, we propose to perform 
all quantum gates solely by means of STIRAP processes.
This technique was originally developed in the field of atomic 
physics~\cite{bergmann:98} as an optimal quantum control strategy 
to channel the system between two long-lived states (here $|0\rangle$ 
and $|1\rangle$) through optical coupling to an interconnecting state 
(here $|3\rangle$): to avoid radiative environment losses of 3, one 
exploits the renormalized radiation-matter states ({\em trapped state}) 
for the transfer process, which is achieved by slowly ({\em 
adiabatically}) varying the exciting laser fields and keeping the 
population of state~3 negligible throughout. 
As a further advantage, such control does not require a detailed 
knowledge of the system parameters (i.e., oscillator strengths) 
and therefore is of very robust nature, thus rendering this scheme 
ideal for quantum control in the solid state~\cite{hohenester.apl:00,brandes:00,pazy:01,hohenester:02}.

{\em Unconditional gates.}\/ Recently, Ki$\check{\rm s}$ and 
Renzoni~\cite{kis:02} extended this original STIRAP level scheme to 
an additional long-lived auxiliary state (here $|2\rangle$), and 
showed that within the resulting model it becomes possible to perform 
generic quantum gates. For the sake of clarity, let us briefly rephrase 
the main steps of this control within the present scheme: suppose 
that initially the system wavefunction is

\begin{equation}\label{eq:initial}
  |\Psi\rangle=
  |L\rangle\otimes
  \left(\alpha|S_x=-\onehalf\rangle+\beta|S_x=+\onehalf\rangle\right).
\end{equation}

\noindent  Next, the quantum dot structure is subject to a first 
STIRAP process, consisting of a sequence of three laser 
pulses: the first one (Stokes pulse) couples the states 2 
and 3; the second ones (pump pulses) affect the 0--3 and 
1--3 transitions, with Rabi frequencies $ \Omega_0(t) = 
\Omega(t)\cos\chi $ and $ \Omega_1(t) = \Omega(t) \exp(i
\eta)\sin\chi$, respectively~\cite{kis:02} (here $\Omega(t)$
is the envelope, and $\chi$ and $\eta$ are phase factors); 
such selective coupling can be achieved by the 
abovementioned selection of the light polarizations and 
frequencies. Incidentally, with a specific choice of the 
laser parameters ($\chi=-\pi/2$ and $\eta=0$) this first 
sequence {\em maps the information stored in the electron 
spin onto the orbital degrees of freedom}:

\begin{equation}\label{eq:processed}
  |\Psi\rangle\longrightarrow
  \left(\alpha|L\rangle+\beta|S\rangle\right)\otimes
|S_x=-\onehalf\rangle :
\end{equation}

\noindent as discussed in the following, this possibility 
is crucial to the implementation of the conditional dynamics
within the present scheme.
The single-qubit rotation is completed by a second, reversed
STIRAP process (pump pulses before the Stokes one): any 
unitary transformation of the $SU_2$  group can be performed
through an appropriate choice of $\chi$ and $\eta$.  
Note that the different 
energies of the involved electron states, Eq.~\eqref{eq:states}, 
result in additional dynamic phase factors, which should be 
incorporated into the quantum algorithm.

{\em Conditional gates.}\/ The conditional (controlled) dynamics can 
be implemented within the present scheme by exploiting the electrostatic 
interaction changes resulting from intermediate population of 2. As an 
illustrative example, let us consider a controlled--NOT gate in a 
structure consisting of two quantum-dot molecules, Fig.~2, where the 
system is initially in state $|1\rangle_c\otimes|0\rangle_t$, with $c$ 
and $t$ denoting the control and target qubit, respectively. At the 
beginning the information of both qubits is encoded in the respective 
electron spins: a first STIRAP process applied to the control qubit 
then maps the $|S_x=+\onehalf\rangle$ component onto the orbital degrees 
of freedom, Fig.~2b, {\em independent of the target qubit setting}.\/ 
In what follows, we shall exploit the fact that this modified charge distribution exerts a potential change on the target qubit and
leads to different transition frequencies. Thus, in the next step, 
Fig.~2c, the double STIRAP pulse sequence discussed above is applied 
to the target qubit with the modified laser frequencies; apparently, 
this procedure rotates the target qubit {\em dependent on the control 
qubit setting}.\/ Finally, the quantum information of the control qubit 
is mapped back to the spin degrees of freedom, Fig.~2d.

By now the strength of our present proposal should have become obvious: 
its ability to map the quantum information between spin and orbital 
degrees of freedom. On the one hand, this allows for a high-finesse 
gating through stimulated Raman adiabatic passage. On the other hand, 
it becomes possible to turn on selectively qubit-qubit interactions 
only during gating; this inter-qubit control emerges naturally for 
the double-dot structure under investigation without requiring 
additional switching of external electric or magnetic fields, and 
appears advantageous over related proposals \cite{biolatti:00,pazy:02} 
where a compromise between optical and interdot coupling had to be 
taken. Thus, the present scheme fully benefits from the long spin 
coherence and the ultrafast optical gating.



{\em 3. Double dot structure.}\/ As a final step, we comment on the possibility to design a quantum dot structure with the desired level scheme of Fig.~1. Quite generally, the relevant features for the implementation of such a scheme are: (i) single-electron wavefunctions sufficiently localized in either dot (to minimize environment losses during gating and to maximize electrostatic potential changes); (ii) a charged-exciton state with the hole delocalized over the double-dot structure (such that all transitions between 0--3, 1--3, and 2--3 aquire comparable oscillator strengths); (iii) energetically well separated transition frequencies $\omega_0$ and $\omega_2$ (in order to energetically resolve the 0--3 and 2--3 transitions, that are induced by optical fields with the same polarization). Model calculations were performed to demonstrate that such manifold requirements can indeed be simultaneously fulfilled. We adopt the framework presented in Refs.~\cite{troiani.prb:02,rontani.ssc:01} where we calculated single- and few-particle states for prototypical GaAs/AlGaAs double-dot structures within the envelope-function and effective-mass approximations, assuming a prototypical confinement potential which is double-well like along $z$ and parabolic in the inplane directions. In addition, we consider: a
slight asymmetry in the double-dot structure (well widths of $l_L=3.5$nm and $l_S=3.6$nm, respectively, and an interdot distance $d=7$nm); an applied electric field; the consideration of a charged-exciton state with light-hole character (to enhance the interdot tunneling of holes; alternatively, it might be advantegeous to use type-II quantum dots where the hole is only Coulomb-bound and its wavefunction becomes strongly delocalized \cite{janssens:02}).

Results of our calculations are shown in Fig. 3. Panel (a) shows the carrier distributions along $z$: the two lower plots represent the single-electron densities $\rho^e_i(z)=|\phi_i^e(z)|^2$; the small overlap between $\phi_{0,1}^e$ and $\phi_2^e$ allows to almost completely suppress environment losses due to phonon-assisted tunneling during gating, as discussed in more detail by Pazy et al. \cite{pazy:01} where lifetimes of the order of nanoseconds where estimated; in the upper part of panel (a) we report the electron (light gray) and hole (dark gray) densities $ \rho^{e,h}_{3}(z) $ of the inter-connecting charged exciton state 3, with $\rho^{e,h}_{3}(z)=\int dx\;dy\; \langle 3|\;\hat{\psi}^{\dagger}_{e,h}({\bm r})
\hat{\psi}_{e,h}({\bm r})\;|3 \rangle$ and $\hat{\psi}_{e,h}({\bm r})$ the field operator for electrons and holes, respectively. While the use of light fields with linear polarizations allows the coupling of the charged exciton with electron states sharing the orbital state and with opposite spins, the present overlap of $\rho^h (z)$ with both $\rho^e_{0,1} (z)$ and $\rho^e_{2} (z)$ ensures comparable oscillator strengths to transitions to states where electrons are localized in opposite dots (irrespective of their spin orientations). This indeed can be seen in Fig.~3(b) where we plot the absorbtion spectra associated to the three 
initial states, with the transitions 0--3 (1--3) and 2--3 indicated by the shaded regions. The additional peaks with larger oscillator strength are attributed to additional intradot transitions; however, the energetic separation between the peaks is of the order of a few meV and is thus certainly large enough to suppress such undesired transitions by use of laser pulse widths of the order of tens of picoseconds (for the polarizability of the single- and few-particle states in artificial molecules see Ref.~\cite{troiani:03}; see also Refs.~\cite{piermarocchi:02,borzi:02} for more sophisticated quantum control strategies). Thus, although more realistic calculations including finer details of the material and dot parameters, e.g., strain distributions or piezoelectric fields \cite{bimberg:98}, might introduce moderate modifications, we believe that our model calculations clearly demonstrate that the level scheme of Fig.~1 can be designed in state-of-the-art quantum dot samples and could open the possibility for much more efficient and sophisticated quantum gates.

 
In conclusion, we have proposed a novel semiconductor-based implementation
scheme for quantum information processing, where the qubit is identified 
with the spin of an excess electron in a vertically coupled double-dot structure.
By use of further auxiliary states it becomes possible to perform all quantum gates by means of stimulated Raman adiabatic passage and to hereby almost completely suppress environment losses during gating. In addition, an efficient mechanism for turning on and off qubit-qubit interactions, as requested for conditional quantum gates, has been proposed. We think that our present work constitutes an important step forward for the implementation of a first proof-of-principle solid-state quantum computer and could open the possibility for highly efficient
quantum gates.

This work has been supported in part by the EU under the IST programme 
``SQID'' and by the Austrian science fund FWF under project P15752.

\newpage

\begin{figure}
\caption{
Schematic representation of our proposed qubit implementation which consists of two vertically coupled quantum dots in presence of an external electric field. (a) Level scheme as described in the text: single-electron states $|0\rangle$ and $|1\rangle$ (electron in large dot $L$ and $|S_x=\mp\onehalf\rangle$, respectively) which are used to encode the quantum information; auxiliary state $|2\rangle$ (electron in small dot $S$ and $|S_x=-\onehalf\rangle$); charged-exciton state $|3\rangle$ which allows optical coupling between all single-electron states through frequency-selective and linearly polarized 
laser pulses (Rabi energies $\Omega_{i,\sigma_i}$, light polarizations $\sigma_{0,2}=x$ and $\sigma_1=y$). (b) [(c)] square modulus of the electron wavefunction along $z$ in the large dot $L$ [small dot $S$].
}
\end{figure}

\begin{figure}
\caption{Implementation of the controlled--NOT gate for an initial state 
$|1\rangle_c\otimes|0\rangle_t$ [panel (a)]; (b) a first adiabatic passage sequence applied to the control qubit transfers the electron to the smaller 
dot $S_c$ and changes the electrostatic potential; (c) a NOT transformation is applied to the target qubit with the laser frequencies tuned to the modified transition energies; (d) a final STIRAP sequence brings back the control qubit to its initial state. The symbols below each dot-molecule indicate the nature of information storage (spin or orbital) at different stages of the quantum gate.}
\end{figure}

\begin{figure}
\caption{
Results of our calculations. (a) Spatial distributions along $z$ for the 
electron states 0, 1, 2 (lower plots) and for the exciton state $X^-$ 
(upper plot; light and dark gray corresponds to electrons and holes, respectively).
The lateral confinement of the carriers is produced by a parabolic potential,
with $\hbar\omega_{e,h}=60,50$~meV, while the applied electric field is 
${\bf F} = -3\hat{\bf z} + 50 (\hat{\bf x} + \hat{\bf y}) $~kV/cm.
(b) Absorbtion spectra correpsonding to the three possible initial states
0, 1 (lower spectrum) and 2 (upper spectrum), where the photon zero  
corresponds to the semiconductor bandgap. 
The shaded regions indicate the energies of the pump and Stokes pulses.
The Zeeman splitting of the 0 and 1 states, induced by the magnetic field 
is neglected in the calculations. The physical parameters (GaAs) are the following: $m^*_{e,h} = 0.067, 0.80 \; m_0$, band offsets $V_0^{e,h} = 400, 215$~meV, dielectric constant $\epsilon =12.9$.}
\end{figure}

\setcounter{figure}{0}
\newpage
\begin{figure}
\centerline{\includegraphics[width=0.85\columnwidth,height=16cm]{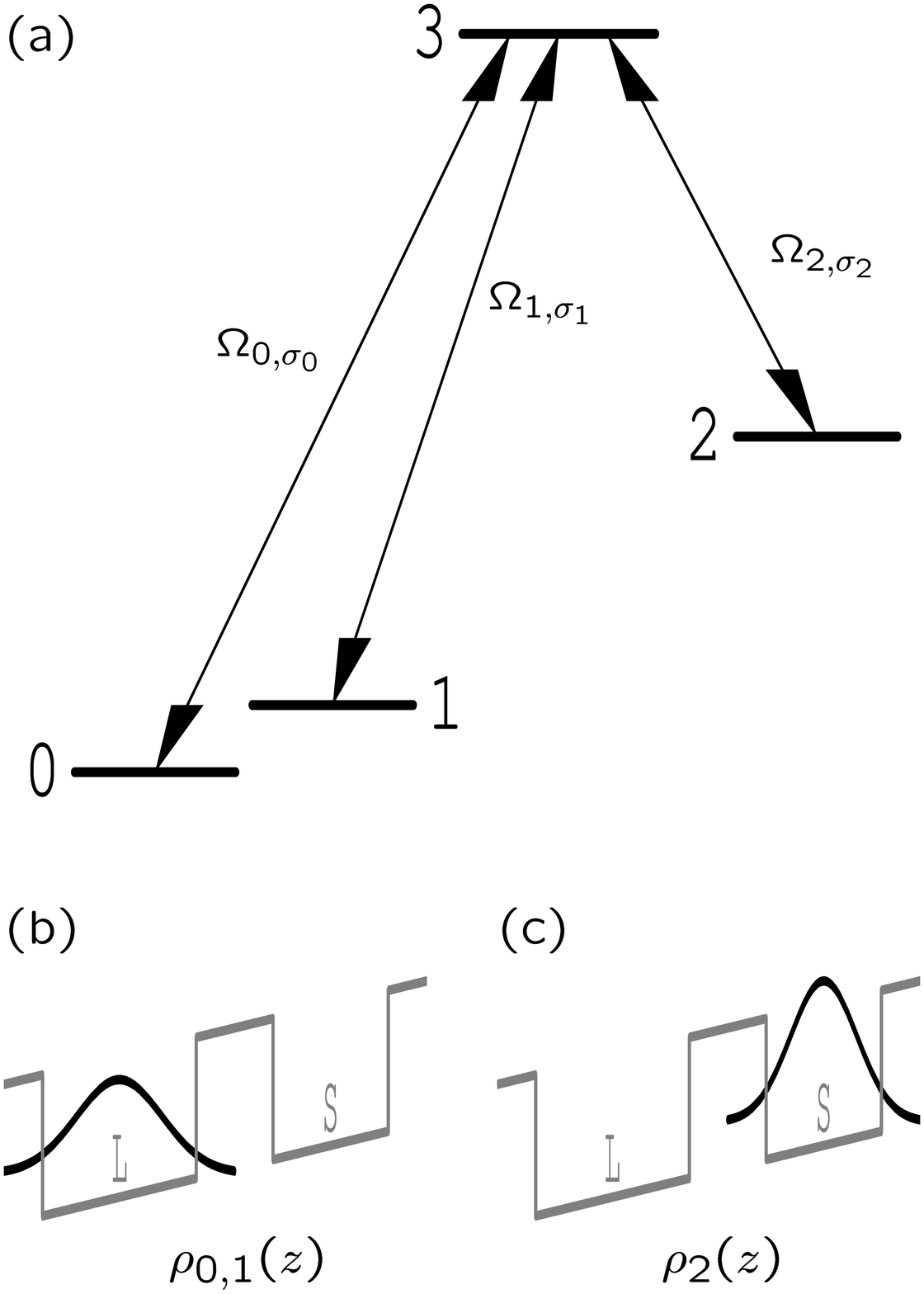}}
\caption{F. Troiani {\em et al.}, High-finesse optical quantum gates for electron spins in artificial molecules}
\end{figure}

\newpage
\begin{figure}
\centerline{\includegraphics[width=0.85\columnwidth,height=16cm]{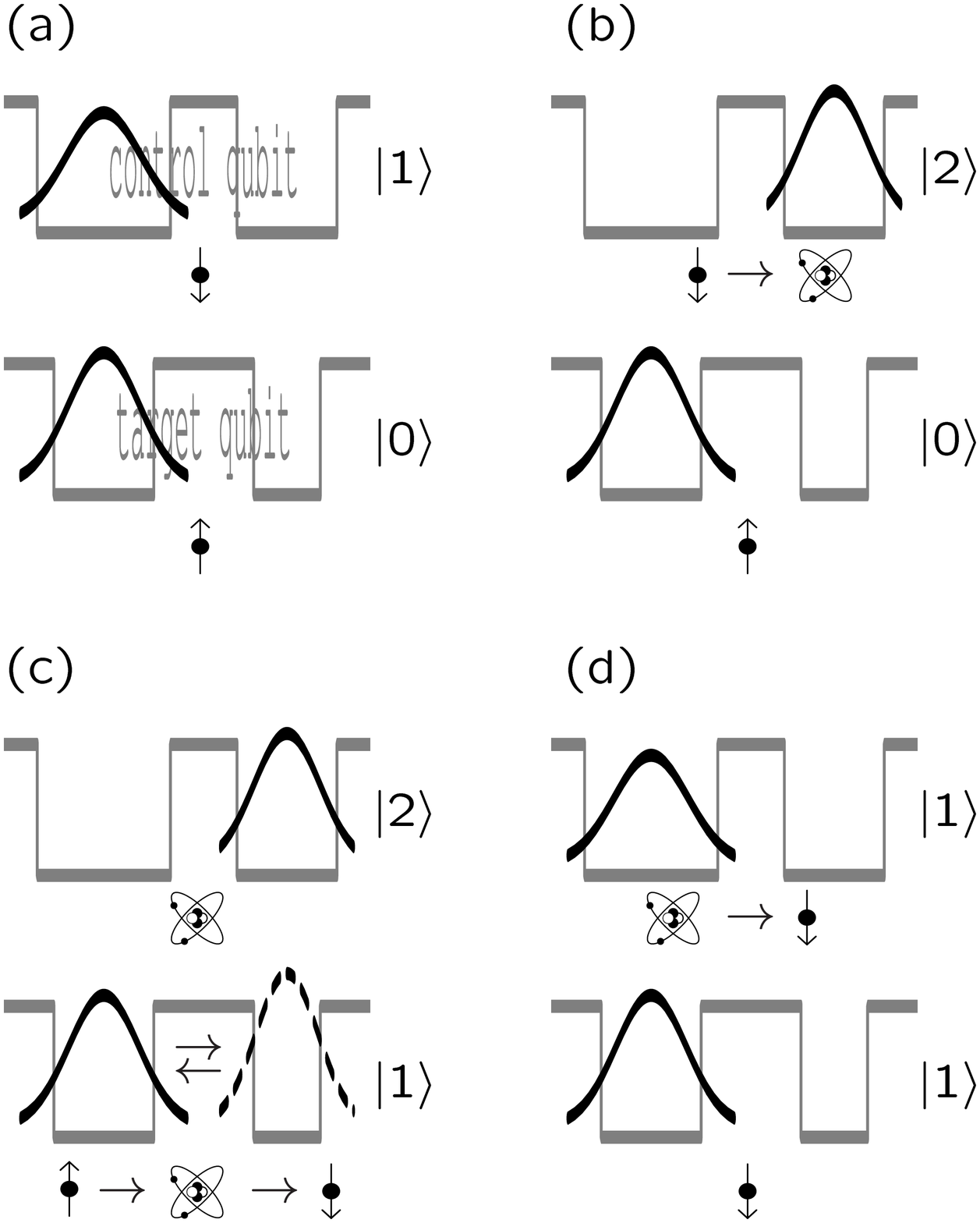}}
\caption{F. Troiani {\em et al.}, High-finesse optical quantum gates for electron spins in artificial molecules}
\end{figure}

\newpage
\begin{figure}
\centerline{\includegraphics[width=0.85\columnwidth,height=20cm]{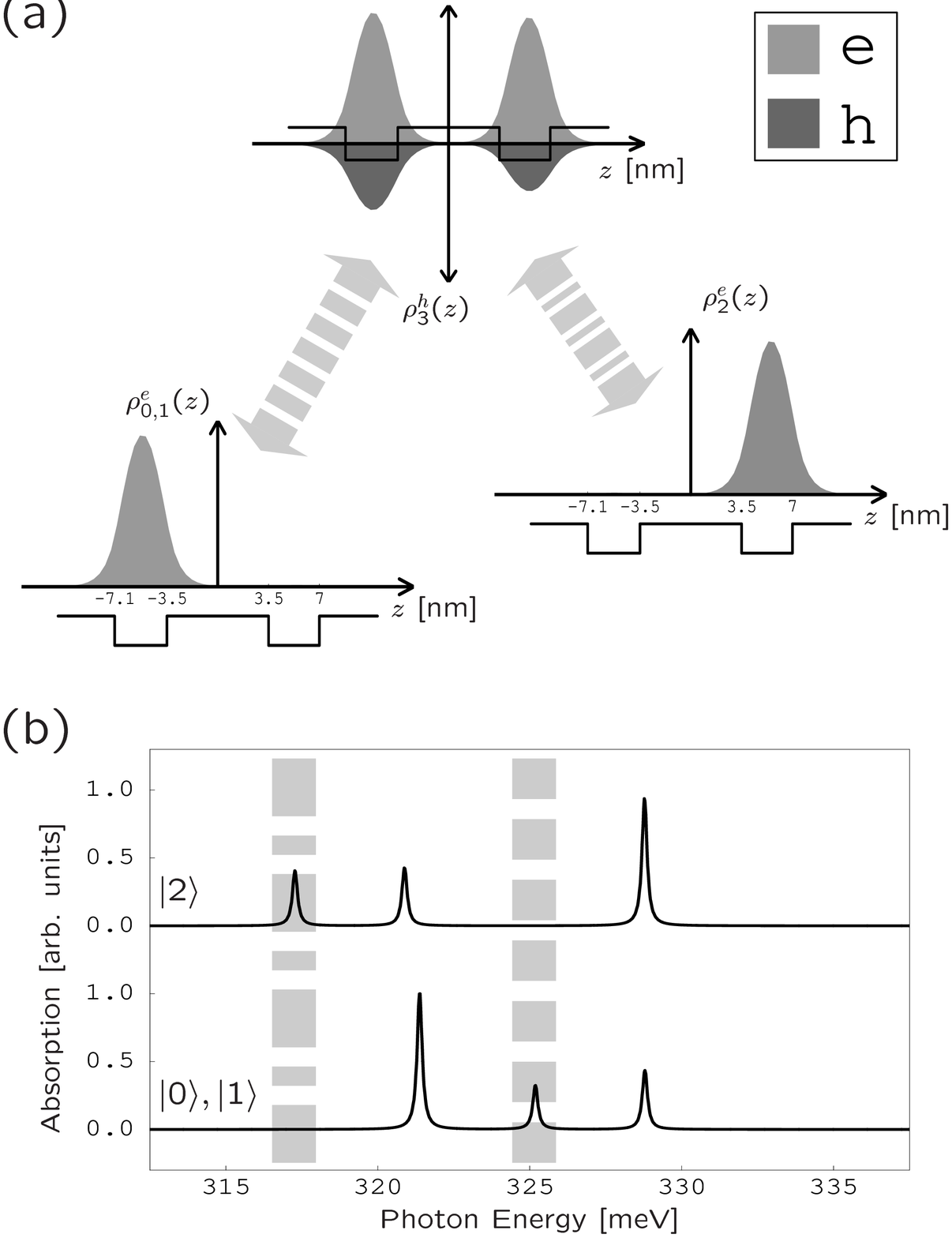}}
\caption{F. Troiani {\em et al.}, High-finesse optical quantum gates for electron spins in artificial molecules}
\end{figure}

\end{document}